\documentclass[twocolumn,showpacs,preprintnumbers,amsmath,amssymb]{revtex4}
\usepackage{graphicx}
\usepackage{dcolumn}
\usepackage{bm}
\begin{document}
\title{Observation of large low frequency resistance fluctuations in metallic  nanowires: Implications on its stability } 
\author{Aveek Bid\footnote[1]{Electronic mail: avik@physics.iisc.ernet.in}$^1$, Achyut Bora$^1$ and A. K. Raychaudhuri\footnote[2]{Electronic mail: arup@bose.res.in}$^{1,2}$}        
\address{$^1$Department of Physics, Indian Institute of Science,  Bangalore 560 012,  India\\
$^2$Unit for Nanoscience and Technology, S.N.Bose National Centre for Basic Sciences, Salt Lake, Kolkata 700098, India}

\begin{abstract}
We have measured the low frequency ($1mHz\leq f \leq 10Hz$) resistance fluctuations in  metallic nanowires (diameter 15nm to 200nm) in the temperature range $77K$ to $400K$. The nanowires were grown electrochemically in polycarbonate membranes and the measurements were carried out in arrays of nanowires by retaining them in the membrane. A large fluctuation in excess of conventional $1/f$ noise  which peaks beyond a certain temperature was found. The fluctuations with a significant low frequency component ($\simeq 1/f^{3/2}$) arise when the diameter of  the wire $\simeq 15nm$ and vanishes rapidly as the diameter is increased.  We argue that Rayleigh-Plateau instability is the likely cause of this excess noise.
\end{abstract}

\maketitle
The  resistance  fluctuation (noise) in  a nanowire is  an important issue both as a problem of basic physics and  as an important input for  the feasibility of using them in nanoelectronic circuits as interconnects. It sets the limit to the best signal to noise ratio one can get in a practical device having such nanowires as components. The thermal noise and the $1/f$ noise  are the  known sources of noise in a metallic system. At very low temperatures shot noise also makes a contribution. The  equilibrium white thermal  noise (the Nyquist Noise) of a wire of resistance $R$ kept at a temperature $T$ can be estimated from the spectral power density $S_{th}\approx 4k_BTR$. However, in a current carrying nanowire a larger contribution (of $ 1/f$ type spectral power density )  is expected to arise from the $\lq\lq$conductance" or $\lq\lq$resistance" noise. There is no {\it apriori} way to estimate the $1/f$ noise, it needs  to be experimentally measured.  In this report we measure low-frequency resistance fluctuations ($1mHz<f<10Hz$) in arrays consisting of Ag and Cu nanowires with diameters down  to  $15nm$ for  $77K \leq T \leq 400K$.  In addition to the usual thermal and $1/f$ noise there are additional sources of resistance noise in metallic nanowires when the diameter is reduced down to $15nm$. These are not present in wires of larger diameter grown by the same method. We find that the source of this extra noise can be traced to  Rayleigh instability ~\cite{r1,r2,chandra} which is expected to occur in systems with large aspect ratio. This extra noise has important implications on  stability of the wire. This instability, if large, can ultimately lead to break down of the wire ~\cite{sem}. To our knowledge, this is the first time that this particular issue has been investigated by measurement of electrical noise in nanowires of this dimension.

\begin{figure}
\begin{center}
\includegraphics[width=7cm]{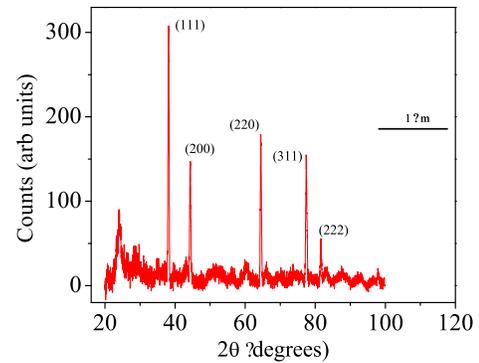}
\end{center}
\vspace{-1cm}
\caption{\label{fig:figure1} XRD of the 15nm Ag nanowires. The inset shows the SEM image of a 200nm wire.}
\end{figure}

The Rayleigh-Plateau instability sets in when the force due to surface tension exceeds the limit that can lead to plastic flow. This occurs at a diameter $ < d_m \approx 2\sigma_S/\sigma_Y$, where $\sigma_S$ is the surface tension and $\sigma_{Y}$ is the Yield force~\cite{r2}. For Ag and Cu (using bulk values for the two quantities) we estimate that $d_m \approx 15nm$. We find that low-frequency resistance fluctuation (noise) measurements, being a very sensitive probe, can observe the instabilities, even though they are small and may not be seen through other measurements. 

\begin{figure}
\begin{center}
\includegraphics{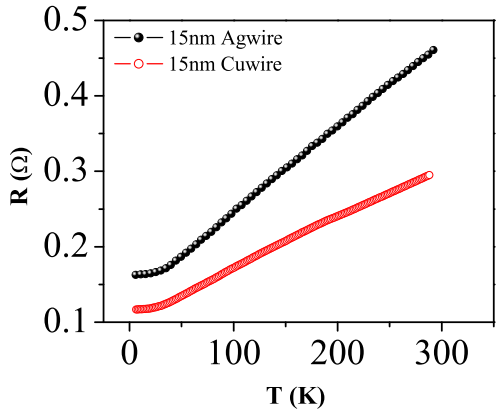}
\end{center}
\vspace{-1cm}
\caption{\label{fig:figure1}Resistance of the 15nm Ag and 15nm Cu samples as a function of temperature. }
\end{figure}

The Ag and Cu  nanowires of approximate length of 6$\mu m$ and diameters 15nm, 20nm and 200nm were grown within polycarbonate templates (etched track membranes)  from AgNO$_3$ and CuSO$_4$~\cite{akr2, chakra} respectively using electrochemical deposition. During the growth one of the electrodes was attached to one side of the membrane and the other electrode was a micro-tip (radius of curvature $\simeq 100 \mu m$) carried by  a micropositioner. This can be placed  at a specific area  on the membrane and thus the growth can be localized. The wires grow by filling the pores from end to end and as soon as one or more wires complete the path from one electrode to the other the growth stops.  The wires can be removed from the membranes by dissolving the polymer in dichloromethane. The wires  were characterized by X-Ray, SEM and TEM. A typical image of a 100nm wire  is shown in figure~1 along with the XRD data. The XRD data matches well with the data taken on a bulk silver sample. We find that typical samples (arrays) on which the measurements were done had 2 to 50 wires. It is very difficult to estimate the exact number of wires that actually grow from end to end. One can use the methods outlined in reference [7]. We find that the estimate made by these methods can be uncertain by a factor of 2. As a result we do not use the exact number of  wires in any of our calculation.

The noise measurement was carried out using a digital signal processing (DSP) based a.c technique (using a lock-in-amplifier) which allows simultaneous measurement of the background noise as well as the bias dependent noise from the sample ~\cite{scoff1,ag1}. The apparatus was calibrated down to a spectral power density $S_V(f)=10^{-20}  V^2/Hz$ by measuring the $4k_{B}TR$ Nyquist noise at a known temperature T of a calibrated resistor.   We have used a transformer preamplifier to couple the sample to the lock-in-amplifier as shown in the inset of figure~2. The carrier frequency was chosen to lie in the eye of the Noise Figure (NF) of the transformer preamplifier to minimize the contribution of the transformer noise to the background noise. The noise appears as the side bands of the carrier frequency.  The  output of the Lock-in amplifier is sampled at a rate of 1024 points/sec by a 16 bit A/D card  and stored in the computer. This forms the time series (consisting of nearly 3 million points) from which the spectral power  density $S_V(f)$ is obtained by using FFT.  The relative variance of the resistance fluctuation $\frac{<(\Delta R)^2>}{R^2}$ within the detection bandwidth ($f_{min}\rightarrow f_{max}$) was obtained by integrating the power spectrum $\equiv \frac{1}{V^2}\int_{f_{min}}^{f_{max}}S_V(f)df$. The lower frequency limit of $1 mHz$ is determined by the quality of the temperature control which is within $\pm 40 ppm$. The output low-pass filter of the Lock-in amplifier had been set at 3 msec with a roll off of  24 dB/octave which has a flat response for $f \leq 10Hz$. This determines the upper limit of our spectral range. 

\begin{figure}
\begin{center}
\includegraphics{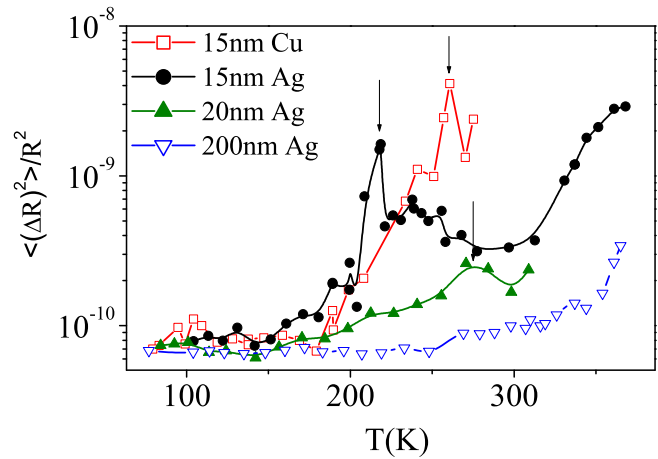}
\end{center}
\vspace{-1cm}
\caption{\label{fig:figure1} The relative variance of the resistance fluctuation  $\frac{<(\Delta R)^2>}{R^2}$ as a function of temperature for the 15nm, 20nm, 200nm  Ag and 15nm Cu nanowires.}
\end{figure}

The resistivity and noise measurements were carried out by retaining the wires within the membrane and the contact were made with silver epoxy. Though the measurements were made with the wires retained within the membrane,  the individual wires are well separated by the insulating walls of the pores in which they were grown. Each wire thus acts as an independent source of noise.The temperature variation for the resistivity measurements were done down to $5K$ in a bath type liquid helium cryostat and the noise was measured down to liquid nitrogen temperature in a similar cryostat.

We note that contact can be a serious issue in noise measurements and we carried out a number of checks to rule out any predominant contribution of the contact to the measured noise. In addition to silver epoxy contacts we also used  evaporated silver films and Pb-Sn solder to make contact.  We find that they give similar results to within $\pm 15\%$. In case of the Pb-Sn contact the change in $R$ of the sample as we go below the superconducting transition temperature ($\sim 7K$) of the solder is negligibly small ($<2-3\%$)implying a small contribution of the contact to the total $R$ measured.  We have also measured the noise in different samples with varying numbers  of wires in them. The resistance of the array, due to different number of wires in them vary.  Also such samples will have significantly different contact areas and hence different contact noise, if any. Yet we find that the normalized noise $\frac{<(\Delta R)^2>}{R^2}$ in different arrays of the same diameter wire lie within $\pm 15\%$.  All these tests rule out any predominant contribution from the contacts. 

\begin{figure}
\begin{center}
\includegraphics{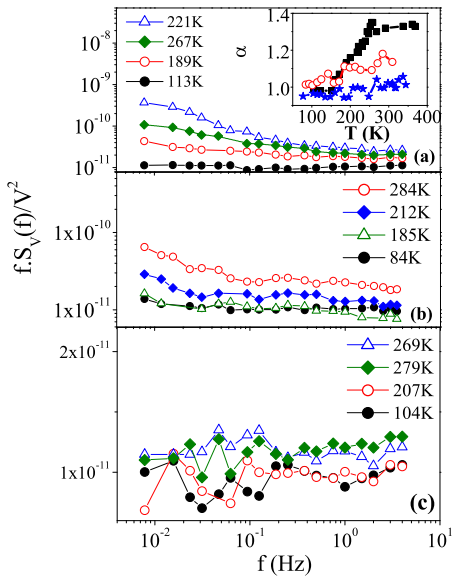}
\end{center}
\vspace{-1cm}
\caption{\label{fig:figure1}  The spectral power density (after subtracting out the measured background noise and multiplying by the frequency) as a function of $f$ at a few representative temperatures for the (a)15nm wire, (b)20nm wire and (c)200nm wire. We have plotted the  spectral power density as  $f.S_V(f)/V^2$ in order to accentuate the deviation from the $1/f$ dependence. The inset of the 4(a) shows the temperature dependence of the measured $\alpha$ for the wires. (filled square-15nm, open circle-20nm and filled star-200nm wire) }
\end{figure}

The 15nm Ag and Cu nanowire arrays have a fairly linear temperature dependence of resistance down to 75K and $R$ reaches a residual value below 50K with  residual resistivity  ratio (RRR)$\frac{\rho_{300K}}{\rho_{4.2K}}  \sim 3$  for both the $15nm$ wires (shown in figure~2).  If the mean free path $l_{e}$ is significantly small due to disorder, then  $k_{F}l_{e}  \approx 1$ ($k_{F}$ = Fermi wavevector) and  the wires would show an  upturn in resistivity at low temperatures due to electron localization. The absence of upturn at low temperature  rules out significant disorder in the system which can give rise to such effects as localization. Another parameter that points to the quality of the wires grown is the temperature coefficient of resistivity $\beta = 1/R(dR/dT)$ as the presence of large degree of structural imperfections and impurity can reduce $\beta$ significantly. For the wires grown $\beta$ lies in the range $\sim 4\times  10^{-3}/K$ and $\sim 2.5\times  10^{-3}/K$  at 300K. For high purity bulk Ag  and Cu  $\beta \simeq3.8\times  10^{-3}/K$ and $3.9\times  10^{-3}/K$ respectively ~\cite{CRC} at 300K.
  
In figure~3 we show a plot of  $\frac{<(\Delta R)^2>}{R^2}$  as a function of T. The spectral power  density $S_V(f)$ is $\propto V^{2}$. The figure clearly shows the enhanced fluctuations in the 15nm wires occuring for $T > 200K$ in comparison to wires having a larger diameter. Absence of a large noise in the $200nm$ wires  establish that the observed features are intrinsic to the $15nm$  wire and that they do not arise due to the presence of the polycarbonate membrane.   In both the 15nm Ag and the Cu nanowires, $\frac{<(\Delta R)^2>}{R^2}$ increases as T is raised and  shows a prominent peak at $T = T^{*}$ (indicated by arrows in figure~3), where $T^{*}$ is $\approx 220K$ for the Ag nanowire and  $\approx 260K$ for the Cu wire. In the $20nm$ wire the fluctuations, though distinctly smaller than that seen in $15nm$ wire, show a shallow peak at $T^* \simeq 275K$. The observed behavior are reproducible and for different samples the variablity in the data is within experimental errors of about $15\%$. For $T>T^*$, $\frac{<(\Delta R)^2>}{R^2}$ shows a shallow shoulder before beginning to rise again. The rise in the fluctuation beyond $300K$ are qualitatively similar in all the samples. The rise is observed even in films of thickness $\simeq 100-200nm$ ~\cite{kar1}.

In figure~4(a) we show $f.S_V(f)/V^2$  as a function of $f$ at a few representative temperatures for the 15nm Ag wire.  The  $S_V(f)$ follows the relation $S_{V}(f) \propto 1/f^{\alpha}$.  The data have been  plotted as   $f.S_V(f)/V^2$ in order to accentuate the deviation of $\alpha$ from 1. The temperature variation of $\alpha$ for the sample is shown as an inset. Similar data were also obtained for the Cu wires. At $T<T^*$, $\alpha \approx 1$ while for $T\geq T^{*}$ we find deviation that $\alpha$ deviates from unity.  For the 15nm wire $\alpha$ increases gradually to $\approx 1.35-1.4$ at $T^*$ and stays at that level till the highest temperature measured. For the 20nm Ag wire also $\alpha \approx 1$ for $T<200K$ and increases to a value  $\alpha \approx 1.2$ at $T^*$, while for  the 200nm wire $\alpha \approx 1$ for the complete temperature range covered. One thus sees that the spectral decomposition of the large noise present in the 15nm wire is qualitatively different from that present in the 200nm wire. The fluctuations in $15nm$ wires  are not only large but also have a large low  frequency component.  

The results indicate  that the nanowires of diameter 15nm have  significantly different resistance fluctuations. There is a  jump in $\frac{<(\Delta R)^2>}{R^2}$ at a certain temperature (which we denote by  $T^*$) and  $S_V(f)$ deviates from  the usual $1/f$ dependence for $T>T^{*}$. The nature of the $1/f$ noise in the 200nm wire is qualitatively similar to that of a  Ag film and this we think is the usual $1/f$ noise in the metallic conductors which can generally be explained by the Dutta-Horn model which gives $\alpha \approx 1$~\cite{Dutta}. The extra fluctuation observed in the 15nm wires at the temperature range around $T^*$ and above is thus expected to arise from a different origin. 

\begin{figure}
\begin{center}
\includegraphics[width=7cm]{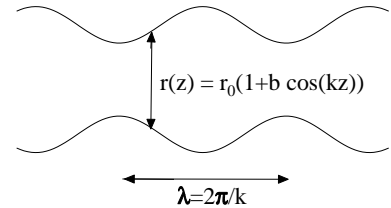}
\end{center}
\vspace{-1cm}
\caption{\label{fig:figure1} Schematic of the model used to estimate the resistance fluctuation arising due to a volume conserving fluctuation in the radius of a wire of finite length.}
\end{figure}

We propose that the extra fluctuation is a manifestation of the Rayleigh-Plateau  instability. The critical diameter $d_m$ at which instability sets in for both Cu and Ag  is about 15nm. To quantitatively estimate the extent of the fluctuation, we make a model shown in figure~5, where the wire has a volume conserving fluctuation of wave vector $k$ in the z-direction. 
The radius of the wire at a distance $z$ along its axis will be given by 
\begin{equation}
r(z)=r_0[1+b~Cos(kz)]
\end{equation}
where the perturbation amplitude $b\ll 1$ and  $r_{0}$ is the radius of the unperturbed wire.
For a volume preserving fluctuation $k$ and $b$ will be related by the relation:
\begin {equation}
\label{b}
b = -\frac{8Sin(kl)}{2kl+Sin(2kl)}
\end{equation}
where $l$ is the length of the wire. The fluctuation in radius will lead to a fluctuation in the cross-section of the wire which in turn will lead to a fluctuation in the resistance which the measurement picks up. We can calculate the resistance $R$ of a wire of length $l$ whose radius $r(z)$ follows relation [1] and [2] and this is given by:
\begin {equation}
\frac{R}{R_0} = \frac{2}{k'(1-b)^3}[\frac{1}{a^3}tan^{-1}(\frac{1}{a}tan\frac{k'}{2})-(\frac{b}{a^2})\frac{tan(k'/2)}{a^2+tan^2(k'/2)}]
\label{rayleigh}
\end{equation}
where $a =\sqrt{(1+b)/(1-b)}$ and $k' = kl$. $R_0$ is the resistance of the unperturbed wire of radius $r_0$. We note that following the Rayleigh criterion~\cite{chandra} for the stability the perturbing modes become unstable when $kr_0 < 1$ and the fastest growing unstable mode occurs for $kr_0 = 0.697$. Using these numbers we can an estimate of the resistance fluctuation from equation ~\ref{rayleigh} . We get an estimate of  $\frac{<(\Delta R)^2>}{R^2} \simeq 10^{-8}$ for the 15nm wire Ag wire at 300K and this compares rather well with  $3\times 10^{-9}$ observed experimentally within the band-width of our experiment. Our simple estimate thus suggests that the extra noise could be linked to the Rayleigh -Plateau instability. The fact that the extra noise appears when the diameter $\approx d_m$ and rapidly goes down even for the $20nm$ wire where the diameter $\approx 1.3d_m$, strongly points to this mechanism for the extra noise. 

The instability of the wire which  leads to long wavelength fluctuations in the radius of the wire may either lead to a break down of the wire into smaller droplets or  can lead to long lived fluctuations without droplet formation.  We believe that the second case holds here because the wire diameter ($15nm$) is the same as the critical diameter $d_m$. It appears that the electronic contribution to the surface energy can lead to a reduction of the fluctuation arising from the Rayleigh instability~\cite{r1}. Theoretically it has been found that due to the electronic contribution the wire (depending on the factor $k_{F}r_{0}$) can have temperature ranges where it is stable and yet can have islands of instabilities~\cite{r2}.

The power  spectrum of the extra fluctuation deviates significantly from the conventional $1/f$ noise due to  low frequency components. This low frequency component can extend over a range of frequency. However, what we measure in the experiment is limited by the  the bandwidth of our experiment. The Rayleigh instability is  sustained by surface diffusion which is  driven by a gradient in chemical potential produced by surface deformation.  Diffusion generally contributes  a spectral power with $S_{V}(f)\simeq 1/f^{3/2}$ ~\cite{kar1,threebytwo}. The observation that for $T \geq T^*$  $\alpha \rightarrow 3/2$ would suggest existence of such modes. It is thus likely that the  low frequency dynamics seen in the fluctuation arises from this  diffusion process. 
 
To summarize,  we find that in metallic Ag and Cu  nanowires of diameter $15nm$ there are large low frequency resistance fluctuations beyond a certain temperature.  The fluctuations differ in characteristics from those usually found in wires of larger diameters. The fluctuation is not the usual $1/f$ noise. It is proposed that the extra fluctuation originates from Rayleigh instability. The measured resistance  fluctuation was found to have  a magnitude similar to that estimated from a simple model of a wire exhibiting volume preserving fluctuation. The extra noise has a strong dependence on the wire diameter and it makes a sharp appearance when the diameter of the wire approaches the critical value. Our observation has impact in the context of stability of nanowires at or above 300K. It raises a fundamental issue on the use of nanowires as interconnects in nanoelectronics.

\subsection{Acknowledgments} 
This work is  supported by DST, Govt. of India and DIT, Govt. of India. A. Bid  acknowledges CSIR  for support and A.Bora acknowledges UGC for support.


\begin{thebibliography}{1}
\bibitem{chandra}Chandrasekhar S.,  {\it Hydrodynamic and Hydromagnetic Stability} (New York: Dover) , (1981).
\bibitem{r1} F Kassubek, C A Stafford, Hermann Grabert and Raymond E Goldstein, Nonlinearity, {\bf 14}, 167, (2001).
\bibitem{r2}C.-H. Zhang, F. Kassubek, and C. A. Stafford1, Phys. Rev. B, {\bf 68}, 165414, (2003)
\bibitem{sem}M. E. Toimil Molares, A. G. Balogh, T. W. Cornelius, R. Neumann, and C. Trautmann, Appl. Phys. Letts, {\bf 85}, 5337, (2004).
\bibitem{akr2}A.K.Raychaudhuri {\it The Chemistry of Nanomaterials}(ed. C.N.R Rao, A.Muller and A.K.Cheetham, WILEY-VCH, N.Y 2004) vol 2, page 688.
\bibitem{chakra} S. Bhattacharrya, S. K. Saha and D. Chakravorty,  Appl. Phys. Letts., {\bf 76}, 3896, (2000).
\bibitem{giordano}W.D.Williams and N. Giordano, Phys. Rev B, {\bf 33}, 8146, (1986).
\bibitem{scoff1} J. H. Scofield,  Rev. Sci. Instr.,  {\bf 58}, 985, (1987).
\bibitem{ag1}A.Ghosh, S.Kar, A.Bid and A.K.Raychaudhuri arXiv:cond-mat/0402130 v1 4 Feb 2004.
\bibitem{CRC}Handbook of Chemistry and Physics. 74th Edition. New York: CRC Press, 1993-1994.
\bibitem{Dutta} P.Dutta and P. M. Horn,  Rev. Mod. Phys., {\bf 53}, 497–516 (1981).
\bibitem{kar1}  Swastik Kar and A. K. Raychaudhuri,  Appl. Phys. Letts, {\bf 81}, 5165, (2002).
\bibitem{threebytwo}B. Fourcade and A.–M. S. Tremblay, Phys. Rev B, {\bf 34}, 7802, (1986).
\end{thebibliography}
\end{document}